\documentclass[a4paper]{article}
\usepackage{geometry}
\usepackage{cleveref}

\usepackage{nameref}

\usepackage[figuresright]{rotating}
\usepackage{graphicx}

\usepackage{longtable}
\usepackage{array}
\usepackage{booktabs}

\usepackage{multirow}
\usepackage{makecell}

\usepackage{graphicx}
\usepackage{tensor}
\usepackage{mathtools}
\usepackage{amsmath}
\usepackage{amsxtra}

\usepackage{upgreek}

\usepackage{appendix}

\usepackage{bm}

\usepackage[noblocks]{authblk}

\usepackage[numbers,sort&compress]{natbib}

\usepackage{CJK}
\usepackage{txfonts}
\usepackage{times}
\usepackage{ragged2e}
\usepackage{indentfirst}

\usepackage{verbatim}

\usepackage{ntheorem}

\renewcommand{\raggedright}{\leftskip=0pt \rightskip=0pt plus 0cm}
\usepackage{setspace}
\allowdisplaybreaks[4]
\begin{document}

\title{Greenberger-Horne-Zeilinger-based quantum private comparison \\ protocol with bit-flipping
\footnote{Project supported by the State Key Program 
of National Natural Science of China (Grant No. 61332019), the
Major State Basic Research Development Program of China
(973 Program, Grant No. 2014CB340601), the National Science
Foundation of China (Grant No. 61202386 and No. 61402339), 
and the National Cryptography Development Fund
(No. MMJJ201701304).}}

\author{Zhaoxu Ji, Peiru Fan, Huanguo Zhang, and Houzhen Wang$^\dag$
\\
{\small {Key Laboratory of Aerospace Information Security and Trusted Computing, 
Ministry of Education, School of Cyber Science and Engineering, 
Wuhan University, Wuhan 430072 China\\
$^\dag$whz@whu.edu.cn}}}
\date{}

%-----------------------------------------------------------------------------------------------------

\maketitle

%-----------------------------------------------------------------------------------------------------

\begin{abstract}
By introducing a semi-honest third party (TP),  we propose in this paper a novel QPC protocol using
$(n+1)$-qubit ($n \ge 2$) Greenberger-Horne-Zeilinger (GHZ) states
as information carriers. The parameter $n$ not only determines the number of
qubits contained in a GHZ state, but also determines the probability that
TP can successfully steal the participants' data and the qubit efficiency. 
In the proposed protocol, we do not employ any other quantum technologies
(e.g., entanglement swapping and unitary operation)
except necessary technologies such as preparing quantum states and quantum measurements,
which can reduce the need for quantum devices.
We use the keys generated by quantum key distribution and bit-flipping for privacy protection,
and decoy photons for eavesdropping checking, making both external and internal attacks invalid.
Specifically, for external attacks, we take several well-known attack means 
(e.g., the intercept-resend attack and the measurement-resend attack) as examples to show 
that the attackers outside the protocol can not steal the participants' data successfully,
in which we provide the security proof of the protocol against the entanglement-measurement attack.
For internal attacks, we show that TP cannot steal the participants' data and 
the participants cannot steal each other's data.
We also show that the existing attack means against QPC protocols are invalid for our protocol.
\end{abstract}

\begin{comment}
In terms of the security of the protocol, w
making both outsider and insider attacks invalid.
and decoy photons for eavesdropping checking.

Furthermore, the GHZ states are prepared
by participants rather than by the third party,
which can reduce potential security risks.
\end{comment}

%-----------------------------------------------------------------------------------------------------



%-----------------------------------------------------------------------------------------------------

\section{Introduction}

%Judging whether two or more data are the same by comparison has a wide range of
%applications in information science
Comparing two or more data to determine whether they are the same 
has a wide range of applications in information science, 
such as malware detection and clustering,
patch generation and analysis, and bug search \cite{ChoiSS812010,VinodP2009}.
A natural question is how to complete the comparison if all the data are confidential. 
This problem is called ``Tierc\'e problem'' or ``socialist millionaires' problem'',
which is originated from the ``millionaires' problem'' raised by Yao in 1982
\cite{ZhangHG16102019,YangYG422009,JiaHY2841,JiZX1249112019}.
The solutions to this problem can also solve many problems in real life,
such as secret bidding and auctions, secret ballot
elections, e-commerce, data mining, and authentication 
\cite{ZhangHG16102019,YangYG422009,JiaHY2841,JiZX1249112019}.

%绪论中介绍QSMC

Quantum private comparison (QPC), as an important branch of quantum secure
multi-party computation (QSMC) \cite{ZhangHG16102019,ZXJi101007}, is the generalization of the solutions 
to the ``Tierc\'e problem'' in quantum mechanics.
The difference between QPC and the classical solutions is that 
its security is based on the principles of quantum mechanics 
rather than computational complexity
\cite{ZhangHG16102019,YangYG422009,JiaHY2841,JiZX1249112019,
ZhangHG58112015,ZXJi101007,JiZX19112019}.
QPC has attracted wide attention from academia in recent decade 
because it can provide unconditional security for real-life information transactions.
A QPC protocol needs to introduce a semi-honest third party (conventionally called TP), 
who faithfully executes protocol processes to
assist participants in completing private comparison
and will not collude with any participant, but he cannot
steal the participants' data in all possible ways \cite{ZhangHG16102019,JiZX1249112019}.
In addition, a QPC protocol should satisfy two conditions \cite{ZhangHG16102019,JiZX1249112019}:
1) fairness: all participants get the comparison result at the same time without a specific order; 
2) security: each participant's data is confidential,
and the other participants, TP and the attackers outside the protocol
cannot successfully steal the participant's data;
iff all participants' data are the same,
the participants know each other's secret data.

%本文考虑如何减少密钥消耗，同时又保证安全
In this paper, we propose a novel QPC protocol.
We use Greenberger-Horne-Zeilinger (GHZ) states as information carriers,
and introduce a semi-honest TP who
assists the participants in completing the protocol without colluding with them.
Unlike most existing protocols, our protocol only uses
single-particle measurement technology instead of
entanglement measurement, entanglement swapping, unitary operation
and other technologies, which naturally reduces the need for expensive quantum devices.
As for the security of the protocol, we make use of the entanglement correlation of the GHZ states, 
and use the techniques of quantum key distribution (QKD)
\cite{AlleaumeR5602014,RennerR6012008,GisinN7412002,
DengFG7012004,LinS8732013,GuoY8142010},
decoy photons and bit-flipping to protect data privacy.
In addition, the GHZ states are prepared by participants rather than by TP,
which can to some extent improve the security and efficiency of the protocol.
We prove in detail that our protocol can resist both external and internal attacks.
In particular, we give the security proof against the entanglement-measurement attack,
which may be of guiding significance to the design and security analysis of other QSMC protocols.

%which is not only applicable to our protocol, but also applicable to other QSMC protocols %绪论中介绍QSMC
%which use GHZ states as information carriers and decoy photon technology for eavesdropping checking.

We arrange the rest of this paper as follows.
In Sec. 2, we first introduce the GHZ states
used in our protocol, and then describe protocol steps in detail.
Sec. 3 is devoted to analyze the protocol security,
including the outsider attack and insider attack.
Sec. 4 gives some useful discussions.
We make a summary in Sec. 5.

\section{Proposed quantum private comparison protocol}

Let us now introduce the GHZ states.
The canonical orthonormal $m$-qubit ($m \in \rm N_+$ and $m \ge 3$) GHZ states,
as information carriers in our protocol, can be expressed as
\begin{align}
\label{GHZ1}
\left|G_{k}^{\pm}\right\rangle = \frac 1{\sqrt{2}}
\Big( \left|B\big(k\big)\right\rangle \pm \left|B\big(2^{m}-k-1 \big)\right\rangle \Big),
\end{align}
where $k = 0,1,\ldots,2^{m-1}-1$, and $B(k) = 0b_2b_3\cdots b_m$ is the
binary representation of $k$ in an $m$-bit string,
thus $k=0\cdot2^{m-1}+b_2\cdot2^{m-2}+b_3\cdot2^{m-3}+\cdots+b_m\cdot2^0$.
Obviously, they are orthonormal and and complete,
\begin{align}
\langle G_{k}^{\pm}|G_{k^{\prime}}^{\pm} \rangle = \delta_{k,k^{\prime}}.
\end{align}
Eq. \ref{GHZ1} can be written in a more concise form as follows:
\begin{align}
\label{GHZ2}
\left|G_{k}^{\pm}\right\rangle = \frac 1{\sqrt{2}}
\Big( \left|0b_2b_3\cdots b_m\right\rangle \pm
\left|1\bar{b}_2\bar{b}_3\cdots\bar{b}_m\right\rangle \Big),
\end{align}
where a bar over a bit value indicates its logical negation.

\subsection{Prerequisites}

Next, let us introduce three prerequisites for the proposed protocol.

\begin{enumerate}

\item Suppose that two participants, Alice and Bob,
have secret data $X$ and $Y$ respectively; the binary representations of
$X$ are $\left(  x_{1},x_{2},\ldots,x_{N} \right)$, and $Y$
$\left(  y_{1}, y_{2},\ldots,y_{N}  \right)$, where
$ x_{j}, y_{j} \in \{ 0, 1 \} \phantom{1} \forall j=1,2,\ldots,N,
X=\sum_{j=1}^{N} x_j2^{j-1},Y=\sum_{j=1}^{N} y_j2^{j-1}$
($N \in \rm N_+$, and $N$ is usually a large number,
just like the value of the millionaire's wealth mentioned in the Introduction).
With the help of the semi-honest third party (TP) who
may behave badly but will not collude with either participant,
Alice and Bob want to judge whether $X = Y$.

\item Alice(Bob) divides the binary representation of $ X(Y) $ into $\lceil N/n \rceil$ groups,
\begin{equation}
G_a^1, G_a^2,\ldots, G_a^{ \lceil \frac N n \rceil } 
( G_b^1, G_b^2,\ldots, G_b^{\lceil \frac N n \rceil}),
\end{equation}
where $n \in \rm N_+$ and $2 \le n \le N$ throughout this protocol,
and each group $ G_a^i (G_b^i) $ includes $n$ bits
($i = 1,2,\ldots,\lceil N/n \rceil$ throughout this protocol).
If $ N $ mod $ n = l $, Alice(Bob) adds $ l $ 0 into the last group
$ G_a^{ \lceil N/n \rceil}$ $( G_b^{\lceil N/n  \rceil} ) $.

\item Alice, Bob and TP agree on the following coding rules:
$\left|0\right\rangle \leftrightarrow 0$ and $\left|1\right\rangle \leftrightarrow 1$;
the coding rules are public.

\end{enumerate}

\subsection{Protocol steps}

Let us now describe in detail the steps of the protocol:
(the flow chart of the protocol is shown in fig. 1):

\begin{enumerate}

\item Step 1: key generation

Alice and Bob use QKD to generate the shared secret key
sequence $\{K_{AB}^1, K_{AB}^2,\ldots,K_{AB}^{\lceil N/n \rceil}\}$
(note that if Alice and Bob are in the same place, 
they can directly generate the shared key sequence without using QKD).
Similarly, Alice and TP generate the shared key sequence
\{$K_{AC}^1$, $K_{AC}^2$,$\ldots$, $K_{AC}^{\lceil N/n \rceil}\}$.
Bob and TP generate the shared key sequence
\{$K_{BC}^1$, $K_{BC}^2$,$\ldots$, $K_{BC}^{\lceil N/n \rceil}\}$.
Here, $K_{AB}^i,K_{AC}^i, K_{BC}^i\in \{0,1\}$,
and note that the keys generated by QKD are confidential 
and are always assumed to be secure in QPC.
Otherwise, the security of the protocol can not be guaranteed and 
the design of the protocol can not be completed.

\item Step 2: encryption

Alice(Bob) computes $K_{AB}^i \oplus K_{AC}^i (K_{AB}^i \oplus K_{BC}^i)$ and 
denotes the computing results as $R_A^i (R_B^i)$
(i.e. $R_A^i = K_{AB}^i \oplus K_{BC}^i, R_B^i = K_{AB}^i \oplus K_{BC}^i$),
where the symbol $\oplus$ denotes the module 2
operation (i.e. XOR operator) throughout this paper.
Then, Alice(Bob) encrypts her(his) data $G_a^i(G_b^i)$
according to the value of $R_A^i (R_B^i)$. Concretely, if $R_A^i = 1 (R_B^i= 1)$,
she(he) flips each bit in $G_a^i(G_b^i)$ (e.g. $0101\rightarrow1010$),
otherwise keeps $G_a^i(G_b^i)$ unchanged.
Finally, Alice(Bob) denotes her(his) encrypted data 
as $G_a^{i'}(G_b^{i'})$, that is,
\begin{align}
G_a^{i'}(G_b^{i'})
=&\begin{cases}
G_a^i(G_b^i), & \text{if } R_A^i = 0 (R_B^i= 0);	\\
\sptilde{G_a^i}(\sptilde{G_b^i}), & \text{if } R_A^i = 1 (R_B^i= 1),
\end{cases}
\end{align}
where the symbol $\sim$ is a bitwise inverse operator.
For example, if $G_a^i = 01101$, then $\sptilde{G_a^i} = 10010$.

\item Step 3: state preparation

According to the value of $G_a^{i'}(G_b^{i'})$,
Alice(Bob) prepares the $(n+1)$-qubit GHZ state
\begin{align}
&\left|G(a_i^0,a_i^1,\ldots,a_i^n)\right\rangle = 
\frac 1{\sqrt{2}}
\big( \left|0a_1a_2\cdots a_n\right\rangle +
\left|1\bar{a}_1\bar{a}_2\cdots\bar{a}_n\right\rangle \big)	\notag \\
\Big(
&\left|G(b_i^0,b_i^1,\ldots,b_i^n)\right\rangle = 
\frac 1{\sqrt{2}}
\big( \left|0b_1b_2\cdots b_n\right\rangle +
\left|1\bar{b}_1\bar{b}_2\cdots\bar{b}_n\right\rangle \big)
\Big),
\end{align}
where $a_1a_2\cdots a_n (b_1b_2\cdots b_n)$ is the
binary representation of $G_a^{i'}(G_b^{i'})$, hence
\begin{align}
&G_a^{i'}=a_1\cdot2^{n-1}+a_2\cdot2^{n-2}+\cdots+a_n\cdot2^0 \notag \\
\big(
&G_b^{i'}=b_1\cdot2^{n-1}+b_2\cdot2^{n-2}+\cdots+b_n\cdot2^0
\big).
\end{align}

Subsequently, Alice(Bob) takes all the particles out from
$\left|G(a_i^0,a_i^1,\ldots,a_i^n)\right\rangle$($\left|G(b_i^0,b_i^1,\ldots,b_i^n)\right\rangle$)
to construct the sequence
\begin{align}
&a_1^0,a_1^1,\ldots,a_1^n, a_2^0,a_2^1,\ldots,a_2^n,
\ldots,a_{\lceil N/n \rceil}^0,a_{\lceil N/n \rceil}^1,\ldots,a_{\lceil N/n \rceil}^n 	\notag\\
\big(
&b_1^0,b_1^1,\ldots,b_1^n, b_2^0,b_2^1,\ldots,b_2^n,
\ldots,b_{\lceil N/n \rceil}^0,b_{\lceil N/n \rceil}^1,\ldots,b_{\lceil N/n \rceil}^n
\big),
\end{align}
and denotes it as $S_a(S_b)$.

\item Step 4: transmission

Alice(Bob) prepares a set of decoy
photons, where each decoy photon is randomly chosen from the four states
$\left|0\right\rangle, \left|1\right\rangle, \left|+\right\rangle, \left|-\right\rangle$
($ \left|\pm\right\rangle $ = $1/\sqrt{2}$ $( \left|0\right\rangle\pm\left|1\right\rangle) $).
Subsequently, Alice(Bob) inserts each decoy photon into $S_a(S_b)$ at a random position;
the new generated sequence is denoted as $S_a^*(S_b^*)$.
Finally, Alice(Bob) sends $S_a^*(S_b^*)$ to TP.

\item Step 5: eavesdropping checking

After confirming TP's receipt of $S_a^*(S_b^*)$,
Alice(Bob) tells TP the positions and bases of the decoy photons.
TP then measures the decoy photons with the bases announced,
and tells Alice(Bob) the measurement outcomes.
Based on the comparison between the initial states and the
measurement outcomes of the decoy photons,
they can judge whether there is eavesdropping in the 
quantum channels. The error rate exceeding the 
predetermined threshold will lead to the 
termination and restart of the protocol, 
otherwise the protocol will proceed to the next step.

\item Step 6: measurement and comparison

TP performs single-particle measurements on
each particle in $S_a$ and $S_b$ with $Z$ basis $\{\left|0\right\rangle, \left|1\right\rangle\}$.
That is, TP measures each particle marked by
$a_i^0,a_i^1,...,a_i^n$($b_i^0,b_i^1,...,b_i^n$) in
$S_a(S_b)$.
Then, according to the coding rules (see the third prerequisite of our protocol), 
TP denotes the binary numbers corresponding to the 
measurement result of the first particle
marked by $a_i^0(b_i^0)$ as $M_{A_i}^1(M_{B_i}^1)$,
and denotes the binary numbers corresponding to
the measurement result of the remaining particles
marked by $a_i^1,...,a_i^n(b_i^1,...,b_i^n)$ as $M_{A_i}^2(M_{B_i}^2)$.

TP computes $M_{A_i}^1 \oplus K_{AC}^i (M_{B_i}^1 \oplus K_{BC}^i)$,
and denotes the computing results as $C_A^i (C_B^i)$.
Then, according to the value of $C_A^i (C_B^i)$,
TP performs the following operations on $M_{A_i}^2 (M_{B_i}^2)$:
If $C_A^i = 1 (C_B^i = 1)$,
TP flips each bit in $M_{A_i}^2(M_{B_i}^2)$,
otherwise keeps $M_{A_i}^2(M_{B_i}^2)$ unchanged.
Denotes the value of $M_{A_i}^2(M_{B_i}^2)$
after the operations as $M_{A_i}^{2'}(M_{B_i}^{2'})$, then
\begin{align}
M_{A_i}^{2'}(M_{B_i}^{2'})
=&\begin{cases}
M_{A_i}^2(M_{B_i}^2), & \text{if } C_A^i = 0 (C_B^i = 0);	\\
\sptilde{M_{A_i}^2}(\sptilde{M_{B_i}^2}), & \text{if } C_A^i = 1 (C_B^i = 1).
\end{cases}
\end{align}
TP computes $M_{A_i}^{2'} \oplus M_{B_i}^{2'}$ and denote the computing results as $R_C^i$.
If $R_C^i = 00\cdots0$ (i.e., the results are all 0),
TP can conclude that $X = Y$, otherwise $X \ne Y$.
Finally, TP publicly announces the comparison result to Alice and Bob.

\end{enumerate}

\begin{figure}[h]
\centerline{\includegraphics[width=5.5in]{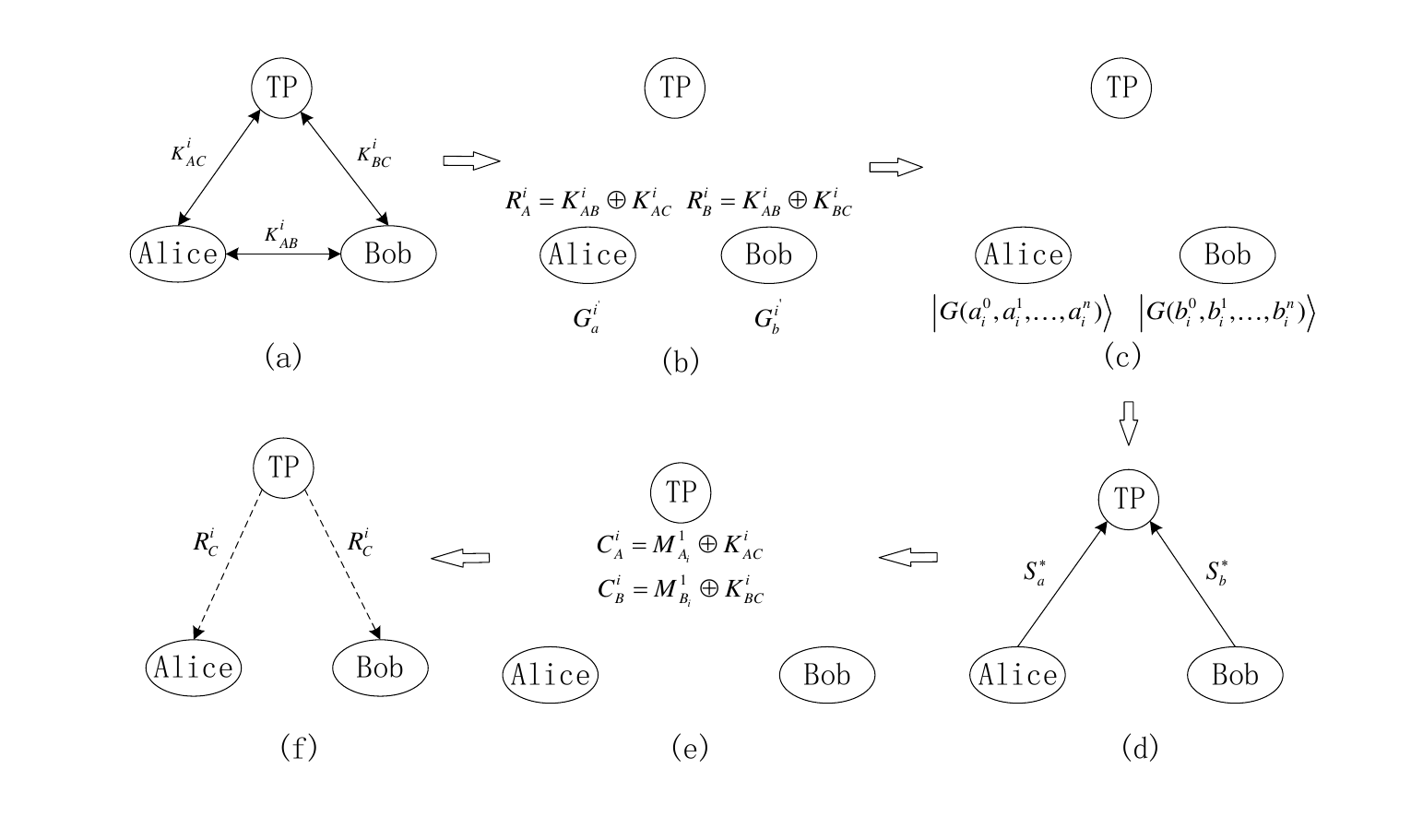}}
\vspace*{8pt}
\caption{The flow chart of the protocol.
The solid arrow lines represent quantum channels, and the
dashed arrow lines classical channels.
For simplicity, we intentionally omit external eavesdroppers and
eavesdropping checking steps.\protect\label{fig1}}
\end{figure}

\subsection{Output correctness}

Let us show that the output of our protocol is correct.
That is, we will show that the value of $M_{A_i}^{2'} \oplus M_{B_i}^{2'}$
equals $G_a^i \oplus G_b^i$. For clarity, we list all possible intermediate 
computational results in Table \ref{truth-table}. As can be seen from the table, 
$M_{A_i}^{2'} \oplus M_{B_i}^{2'}$ is always equal to $G_a^i \oplus G_b^i$ or 
$\sptilde{G_a^i}$ $ \oplus$ $\sptilde{G_b^i}$.
Next let us prove that the equation
$G_a^i \oplus G_b^i$ = $\sptilde{G_a^i}$ $ \oplus$ $\sptilde{G_b^i}$ always holds.
Indeed, we only need to prove that two bits satisfies the equation
because XOR is a bitwise operator.
Suppose there are two bits $p$ and $q$, where $p,q \in \{0,1\}$.
Due to $p \oplus q = (\lnot p \land q) \lor (p \land \lnot q)$, 
$\lnot p \oplus \lnot q = (\lnot (\lnot p) \land \lnot q) \lor (\lnot p \land \lnot (\lnot q))
= (p \land \lnot q) \lor (\lnot p \land q) = p \oplus q $.
Therefore, $G_a^i \oplus G_b^i$ = $\sptilde{G_a^i}$ $ \oplus$ $\sptilde{G_b^i}$.

\begin{comment}

\begin{figure}[t]
\centerline{\includegraphics[width=3.0in]{pic1.png}}
\vspace*{8pt}
\caption{The flow chart of our protocol.
The solid arrow lines represent quantum channels, and the
dashed arrow lines represent classical channels.
For simplicity, we intentionally neglect external eavesdroppers and
eavesdropping checking processes.
\label{fig1}}
\end{figure}

$K_{AB}^i$  &  $K_{AC}^i$ & $K_{BC}^i$ & 
$R_A^i (K_{AB}^i \oplus K_{AC}^i)$ & $R_B^i (K_{AB}^i \oplus K_{BC}^i)$ &
$M_{A_i}^1$ & $M_{B_i}^1$ & $C_A^i(M_{A_i}^1 \oplus K_{AC}^i)$ & 
$C_B^i (M_{B_i}^1 \oplus K_{BC}^i)$ & $M_{A_i}^2$ & $M_{B_i}^2$ \\
\end{comment}

\begin{table}[h]
\begin{spacing}{1.0}
\centering
\setlength{\tabcolsep}{10pt}

\caption{The truth table}
\label{truth-table}
\begin{tabular}{ccccc|cccccc}

\hline\noalign{\smallskip}

$K_{AB}^i$  &  $K_{AC}^i$ & $K_{BC}^i$ & 
$M_{A_i}^1$ & $M_{B_i}^1$ & 
$R_A^i$ & $R_B^i$ &
$C_A^i$ & $C_B^i$ & $M_{A_i}^{2'}$ & $M_{B_i}^{2'}$ \\

\noalign{\smallskip}\hline\noalign{\smallskip}
 
0 & 0 & 0 & 0 & 0 & 0 & 0 & 0 & 0 & $G_a^i$ & $G_b^i$ \\ 

0 & 0 & 0 & 0 & 1 & 0 & 0 & 0 & 1 & $G_a^i$ & $G_b^i$ \\ 

0 & 0 & 0 & 1 & 0 & 0 & 0 & 1 & 0 & $G_a^i$ & $G_b^i$ \\ 

0 & 0 & 0 & 1 & 1 & 0 & 0 & 1 & 1 & $G_a^i$ & $G_b^i$ \\

0 & 0 & 1 & 0 & 0 & 0 & 1 & 0 & 1 & $G_a^i$ & $G_b^i$ \\

0 & 0 & 1 & 0 & 1 & 0 & 1 & 0 & 0 & $G_a^i$ & $G_b^i$ \\

0 & 0 & 1 & 1 & 0 & 0 & 1 & 1 & 1 & $G_a^i$ & $G_b^i$ \\  

0 & 0 & 1 & 1 & 1 & 0 & 1 & 1 & 0 & $G_a^i$ & $G_b^i$ \\

0 & 1 & 0 & 0 & 0 & 1 & 0 & 1 & 0 & $G_a^i$ & $G_b^i$ \\

0 & 1 & 0 & 0 & 1 & 1 & 0 & 1 & 1 & $G_a^i$ & $G_b^i$ \\

0 & 1 & 0 & 1 & 0 & 1 & 0 & 0 & 0 & $G_a^i$ & $G_b^i$ \\

0 & 1 & 0 & 1 & 1 & 1 & 0 & 0 & 1 & $G_a^i$ & $G_b^i$ \\

0 & 1 & 1 & 0 & 0 & 1 & 1 & 1 & 1 & $G_a^i$ & $G_b^i$ \\

0 & 1 & 1 & 0 & 1 & 1 & 1 & 1 & 0 & $G_a^i$ & $G_b^i$ \\

0 & 1 & 1 & 1 & 0 & 1 & 1 & 0 & 1 & $G_a^i$ & $G_b^i$ \\

0 & 1 & 1 & 1 & 1 & 1 & 1 & 0 & 0 & $G_a^i$ & $G_b^i$ \\

1 & 0 & 0 & 0 & 0 & 1 & 1 & 0 & 0 & $\sptilde{G_a^i}$ & $\sptilde{G_b^i}$ \\

1 & 0 & 0 & 0 & 1 & 1 & 1 & 0 & 1 & $\sptilde{G_a^i}$ & $\sptilde{G_b^i}$ \\

1 & 0 & 0 & 1 & 0 & 1 & 1 & 1 & 0 & $\sptilde{G_a^i}$ & $\sptilde{G_b^i}$ \\

1 & 0 & 0 & 1 & 1 & 1 & 1 & 1 & 1 & $\sptilde{G_a^i}$ & $\sptilde{G_b^i}$ \\

1 & 0 & 1 & 0 & 0 & 1 & 0 & 0 & 1 & $\sptilde{G_a^i}$ & $\sptilde{G_b^i}$ \\

1 & 0 & 1 & 0 & 1 & 1 & 0 & 0 & 0 & $\sptilde{G_a^i}$ & $\sptilde{G_b^i}$ \\

1 & 0 & 1 & 1 & 0 & 1 & 0 & 1 & 1 & $\sptilde{G_a^i}$ & $\sptilde{G_b^i}$ \\

1 & 0 & 1 & 1 & 1 & 1 & 0 & 1 & 0 & $\sptilde{G_a^i}$ & $\sptilde{G_b^i}$ \\

1 & 1 & 0 & 0 & 0 & 0 & 1 & 1 & 0 & $\sptilde{G_a^i}$ & $\sptilde{G_b^i}$ \\

1 & 1 & 0 & 0 & 1 & 0 & 1 & 1 & 1 & $\sptilde{G_a^i}$ & $\sptilde{G_b^i}$ \\

1 & 1 & 0 & 1 & 0 & 0 & 1 & 0 & 0 & $\sptilde{G_a^i}$ & $\sptilde{G_b^i}$ \\

1 & 1 & 0 & 1 & 1 & 0 & 1 & 0 & 1 & $\sptilde{G_a^i}$ & $\sptilde{G_b^i}$ \\

1 & 1 & 1 & 0 & 0 & 0 & 0 & 1 & 1 & $\sptilde{G_a^i}$ & $\sptilde{G_b^i}$ \\

1 & 1 & 1 & 0 & 1 & 0 & 0 & 1 & 0 & $\sptilde{G_a^i}$ & $\sptilde{G_b^i}$ \\

1 & 1 & 1 & 1 & 0 & 0 & 0 & 0 & 1 & $\sptilde{G_a^i}$ & $\sptilde{G_b^i}$ \\

1 & 1 & 1 & 1 & 1 & 0 & 0 & 0 & 0 & $\sptilde{G_a^i}$ & $\sptilde{G_b^i}$ \\

\noalign{\smallskip}\hline

\end{tabular}
\end{spacing}
\end{table}

\section{Security analysis}

Let us now analyze the security of the proposed protocol.
We would first like to show the attacks from the eavesdroppers outside the protocol
(i.e., the external attack) are invalid. We then show that the internal attacks, 
including those by participants and TP, are also invalid.

\subsection{External attack}

Generally, the external attack refers to the attempt by 
someone outside the protocol to steal the participants' secret data
from a quantum channel. In our protocol, we use the keys generated by QKD and bit-flipping to encrypt the participants' data,
and decoy photon technology to check the security of the quantum channels.
It is known that the decoy photon technology is derived from QKD, and
has been proved unconditionally safe \cite{RennerR6012008}.
With this technology, Eve will be caught no matter what kind of attack means he uses,
such as the intercept-resend attack, the measurement-resend
attack, the entanglement-measurement attack and the denial-of-service attack \cite{JiZX1249112019,JiZX1652016,YeTY562017}.

In what follows, we would like to analyze the security of our protocol 
against the well-known attacks including the intercept-resend attack, the measurement-resend attack
and the entanglement-measurement attack.
we will first analyze the security of the protocol under the first two attacks,
and then provide the security proof against the last one.
%We will finally briefly analyze the security in the presence of quantum channel noise.

\subsubsection{Security against the intercept-resend attack and the measurement-resend attack}
\label{intercept-resend}

The intercept-resend attack can be described as follows:
An eavesdropper (conventionally called Eve) 
intercepts all the particles sent from TP to the participants in Step 4,
and replaces these particles with fake ones.
Then Eve performs single particle measurements on each particle in 
the $(n+1)$-qubit GHZ states after the participants tells TP the positions and bases of decoy photons.
However, he cannot steal the participant's data because the data is encrypted by bit-flipping and
the keys generated by QKD, and he will be detected by the participants with the probability of $1-(3/4)^l$
when $l$ decoy photons are used for eavesdropping checking in Step 5,
where the probability will get closer and closer to 1 with the increase in $l$.

The measurement-resend attack refers to Eve measures directly all the particles after intercepting them,
in which case he will be easily detected by the participants in eavesdropping checking, 
and he cannot obtain any useful information because he 
cannot distinguish the decoy photons from the particles in $(n+1)$-qubit GHZ states.

\subsubsection{Security against the entanglement-measurement attack}

The entanglement-measurement attack can be described as follows:
Eve intercepts part or all of the particles transmitted
between Alice(Bob) and TP in Step 4,
and entangles them with the ancillary particles that she prepares beforehand,
and then resends them to TP. Finally, Eve performs measurements on the ancillary particles
to extract the information carried by them.

Let us now analyze the case that Eve intercepts the particles sent from Alice to TP
(as for the interception of the particles sent from Bob to TP, 
analysis can be carried out in the same way).
Let us denote Eve's unitary operator as $U$, without loss of generality, 
Eve's entanglement action can be expressed as
\begin{align}
\label{action}
U \left|0\right\rangle \left|\varepsilon\right\rangle = 
\lambda_{00} \left|0\right\rangle \left| \epsilon_{00} \right\rangle 
+ \lambda_{01} \left|1\right\rangle \left| \epsilon_{01} \right\rangle,
U \left|1\right\rangle \left|\varepsilon\right\rangle = 
\lambda_{10} \left|0\right\rangle \left| \epsilon_{10} \right\rangle 
+ \lambda_{11} \left|1\right\rangle \left| \epsilon_{11} \right\rangle,
\end{align}
where $\left| \epsilon_{00} \right\rangle$, $\left| \epsilon_{01} \right\rangle$,
$\left| \epsilon_{10} \right\rangle$, $\left| \epsilon_{11} \right\rangle$
are the pure states determined only by $U$, $\left|\varepsilon\right\rangle$ is an ancillary particle, 
and $||\lambda_{00}||^2+||\lambda_{01}||^2 = 1$, $||\lambda_{10}||^2+||\lambda_{11}||^2 = 1$.

When $U$ is performed on the decoy states $\left|+\right\rangle$ and $\left|-\right\rangle$, 
one can get
\begin{align}
\label{decoy1}
U \left|+\right\rangle \left|\varepsilon\right\rangle & = \frac 1{\sqrt{2}}
( 
\lambda_{00} \left|0\right\rangle \left| \epsilon_{00} \right\rangle 
+ \lambda_{01} \left|1\right\rangle \left| \epsilon_{01} \right\rangle
+\lambda_{10} \left|0\right\rangle \left| \epsilon_{10} \right\rangle 
+ \lambda_{11} \left|1\right\rangle \left| \epsilon_{11} \right\rangle
)		\notag \\
&=\frac 12
\left|+\right\rangle	(
\lambda_{00} \left| \epsilon_{00} \right\rangle 
+ \lambda_{01} \left| \epsilon_{01} \right\rangle
+\lambda_{10} \left| \epsilon_{10} \right\rangle 
+ \lambda_{11} \left| \epsilon_{11} \right\rangle
)		\notag \\
&\quad+\frac 12
\left|-\right\rangle	(
\lambda_{00} \left| \epsilon_{00} \right\rangle 
- \lambda_{01} \left| \epsilon_{01} \right\rangle
+ \lambda_{10} \left| \epsilon_{10} \right\rangle 
- \lambda_{11} \left| \epsilon_{11} \right\rangle
),
\end{align}
and
\begin{align}
\label{decoy2}
U \left|-\right\rangle \left|\varepsilon\right\rangle & = \frac 1{\sqrt{2}}
( 
\lambda_{00} \left|0\right\rangle \left|\epsilon_{00}\right\rangle 
+ \lambda_{01} \left|1\right\rangle \left|\epsilon_{01}\right\rangle 
- \lambda_{10} \left|0\right\rangle \left| \epsilon_{10} \right\rangle 
- \lambda_{11} \left|1\right\rangle \left| \epsilon_{11} \right\rangle
)		\notag \\
&=\frac 12
\left|+\right\rangle	(
\lambda_{00} \left| \epsilon_{00} \right\rangle 
+ \lambda_{01} \left| \epsilon_{01} \right\rangle
-\lambda_{10} \left| \epsilon_{10} \right\rangle 
- \lambda_{11} \left| \epsilon_{11} \right\rangle
)		\notag \\
&\quad+\frac 12
\left|-\right\rangle	(
\lambda_{00} \left| \epsilon_{00} \right\rangle 
- \lambda_{01} \left| \epsilon_{01} \right\rangle
-\lambda_{10} \left| \epsilon_{10} \right\rangle 
+ \lambda_{11} \left| \epsilon_{11} \right\rangle
).
\end{align}
From Eqs. \ref{action}, \ref{decoy1}, and \ref{decoy2},
if Eve wants to avoid introducing errors in the eavesdropping checking step, 
$U$ must meet the following conditions:
\begin{align}
\label{correlation}
\lambda_{01} = \lambda_{10} = 0,
\lambda_{00} \left| \epsilon_{00} \right\rangle = \lambda_{11} \left| \epsilon_{11} \right\rangle.
\end{align}

Next, let us demonstrate that entanglement-measurement attack is invalid to our protocol.
Concretely, Eve entangles the $(n+1)$-qubit GHZ state
$\left|G(a_i^0,a_i^1,\ldots,a_i^n)\right\rangle$ prepared by Alice with ancillary particles,
and then measures the ancillary particles to extract information.
This is how Eve attempts to steal Alice's data,
which will be proved invalid below.

From Eqs. \ref{action}, \ref{decoy1}, \ref{decoy2}, and \ref{correlation}, 
if $U$ acts on the single particle state $\left|a\right\rangle$ where $a \in \{0,1\} $, 
then
\begin{align}
\label{single}
U \left|a\right\rangle \left|\varepsilon\right\rangle = 
&\lambda_{a0} \left|0\right\rangle \left| \epsilon_{a0} \right\rangle 
+ \lambda_{a1} \left|1\right\rangle \left| \epsilon_{a1} \right\rangle		\notag \\
=&\begin{cases}
\lambda_{00} \left|0\right\rangle \left| \epsilon_{00} \right\rangle, & \text{if } a = 0;	\\
\lambda_{11} \left|1\right\rangle \left| \epsilon_{11} \right\rangle, & \text{if } a = 1,
\end{cases}	\notag \\
=&\lambda_{aa} \left|a\right\rangle \left| \epsilon_{aa} \right\rangle,
\end{align}
hence we arrive at
\begin{align}
\label{fomerGHZ}
&U_0 \otimes U_1 \otimes \cdots \otimes U_n \left|0a_1a_2\cdots a_n\right\rangle
\left|\varepsilon\right\rangle_0 \left|\varepsilon\right\rangle_1 \otimes \cdots \otimes
\left|\varepsilon\right\rangle_n  \notag \\
&= U_0 \left|0\right\rangle \left|\varepsilon\right\rangle_0
\otimes U_1 \left|a_1\right\rangle \left|\varepsilon\right\rangle_1
\otimes \cdots \otimes
U_n \left|a_n\right\rangle \left|\varepsilon\right\rangle_n 	\notag \\
%%%%%%%%%%%%%%%%%%%%%%%%%%%%%%%%%%%%%%
& = \lambda_{00} \left|0\right\rangle \left|\epsilon\right\rangle_{00}
\otimes \lambda_{a_1a_1} \left|a_1\right\rangle \left| \epsilon_{a_1a_1} \right\rangle
\otimes \cdots
\otimes
\lambda_{a_na_n} \left|a_n\right\rangle \left| \epsilon_{a_na_n} \right\rangle
\notag	\\
%%%%%%%%%%%%%%%%%%%%%%%%%%%%%%%%%%%%%%
& = \lambda_{00}	\lambda_{a_1a_1} \cdot \cdots \cdot \lambda_{a_na_n}
\left|0a_1a_2\cdots a_n\right\rangle
\left| \epsilon_{00} \right\rangle \left| \epsilon_{a_1a_1} \right\rangle
\otimes \cdots
\otimes
\left| \epsilon_{a_na_n} \right\rangle,
\end{align}
where $U_i (i \in \{0,1,\ldots,n\}$ 
denotes the unitary operator acting on the particle $\left|a_i\right\rangle$, 
and $\left|\varepsilon\right\rangle_i$ denotes the ancillary 
particle entangled on $\left|a_i\right\rangle$. Similarly, we arrive at
\begin{align}
\label{laterGHZ}
&U_0 \otimes U_1 \otimes \cdots \otimes U_n 
\left|1\bar{a}_1\bar{a}_2\cdots\bar{a}_n\right\rangle
\left|\varepsilon\right\rangle_0 \left|\varepsilon\right\rangle_1 \otimes \cdots \otimes
\left|\varepsilon\right\rangle_n  \notag \\
&= U_0 \left|1\right\rangle \left|\varepsilon\right\rangle_0
\otimes U_1 \left|\bar{a}_1\right\rangle \left|\varepsilon\right\rangle_1
\otimes \cdots \otimes
U_n \left|\bar{a}_n\right\rangle \left|\varepsilon\right\rangle_n 	\notag \\
%%%%%%%%%%%%%%%%%%%%%%%%%%%%%%%%%%%%%%
& = \lambda_{00} \left|1\right\rangle \left|\epsilon\right\rangle_{00}
\otimes \lambda_{\bar{a}_1\bar{a}_1} \left|\bar{a}_1\right\rangle 
\left|\epsilon_{\bar{a}_1\bar{a}_1}\right\rangle
\otimes \cdots
\otimes
\lambda_{\bar{a}_n\bar{a}_n} \left|\bar{a}_n\right\rangle 
\left|\epsilon_{\bar{a}_n\bar{a}_n}\right\rangle
\notag	\\
%%%%%%%%%%%%%%%%%%%%%%%%%%%%%%%%%%%%%%
& = \lambda_{00}	\lambda_{a_1a_1} \cdot \cdots \cdot \lambda_{a_na_n}
\left|1\bar{a}_1\bar{a}_2\cdots\bar{a}_n\right\rangle
\left| \epsilon_{00} \right\rangle \left| \epsilon_{a_1a_1} \right\rangle
\otimes \cdots
\otimes
\left| \epsilon_{a_na_n} \right\rangle,
\end{align}
note here that we use
$\lambda_{a_ia_i} \left| \epsilon_{00} \right\rangle = 
\lambda_{\bar{a}_i\bar{a}_i} \left| \epsilon_{\bar{a}_i\bar{a}_i} \right\rangle$
(see Eq. \ref{correlation}) in above equation.
From Eqs. \ref{fomerGHZ} and \ref{laterGHZ}, we arrive at
\begin{align}
\label{GHZ}
&U_0 \otimes U_1 \otimes \cdots \otimes U_n 
\left|G(a_i^0,a_i^1,\ldots,a_i^n)\right\rangle
\left|\varepsilon\right\rangle_0 \left|\varepsilon\right\rangle_1 
\otimes \cdots \otimes
\left|\varepsilon\right\rangle_n  \notag \\
%%%%%%%%%%%%%%%%%%%%%%%%%%%%%%%%%%%%%%
&=U_0 \otimes U_1 \otimes \cdots \otimes U_n  
\frac 1{\sqrt{2}}
\big( \left|0a_1a_2\cdots a_n\right\rangle +
\left|1\bar{a}_1\bar{a}_2\cdots\bar{a}_n\right\rangle \big)
\left|\varepsilon\right\rangle_0 \left|\varepsilon\right\rangle_1 
\otimes \cdots \otimes
\left|\varepsilon\right\rangle_n  \notag \\
%%%%%%%%%%%%%%%%%%%%%%%%%%%%%%%%%%%%%%
&=\frac 1{\sqrt{2}}
\Big(
U_0 \otimes U_1 \otimes \cdots \otimes U_n \left|0a_ia_i \cdots a_i\right\rangle
\left|\varepsilon\right\rangle_0 \left|\varepsilon\right\rangle_1 
\otimes \cdots \otimes
\left|\varepsilon\right\rangle_n  \notag \\
& \qquad + 
U_0 \otimes U_1 \otimes \cdots \otimes U_n 
\left|1\bar{a}_1\bar{a}_2\cdots\bar{a}_n\right\rangle
\left|\varepsilon\right\rangle_0 \left|\varepsilon\right\rangle_1
\otimes \cdots \otimes
\left|\varepsilon\right\rangle_n
\Big) \notag \\
%%%%%%%%%%%%%%%%%%%%%%%%%%%%%%%%%%%%%%
&=\frac 1{\sqrt{2}}
\Big(
\lambda_{00}	\lambda_{a_1a_1} \cdot \cdots \cdot \lambda_{a_na_n}
\left|0a_1a_2\cdots a_n\right\rangle
\left| \epsilon_{00} \right\rangle \left| \epsilon_{a_1a_1} \right\rangle
\otimes \cdots \otimes
\left| \epsilon_{a_na_n} \right\rangle 
\notag \\
& \qquad +
\lambda_{00}	\lambda_{a_1a_1} \cdot \cdots \cdot \lambda_{a_na_n}
\left|1\bar{a}_1\bar{a}_2\cdots\bar{a}_n\right\rangle
\left| \epsilon_{00} \right\rangle \left| \epsilon_{a_1a_1} \right\rangle
\otimes \cdots \otimes
\left| \epsilon_{a_na_n} \right\rangle
\Big)
\notag \\
&=
\lambda_{00}	\lambda_{a_1a_1} \cdot \cdots \cdot \lambda_{a_na_n}
\left|G(a_i^0,a_i^1,\ldots,a_i^n)\right\rangle
\left| \epsilon_{00} \right\rangle \left| \epsilon_{a_1a_1} \right\rangle
\otimes \cdots \otimes
\left| \epsilon_{a_na_n} \right\rangle
\end{align}
From the equation, no error will be introduced iff the
ancillary particles and the intercepted particles are in product states.
Therefore, the entanglement-measurement attack is invalid to our protocol.

The above conclusion can also be obtained by mathematical induction.
From Eqs. \ref{fomerGHZ}, \ref{laterGHZ} and \ref{GHZ}, 
it is only necessary to prove that Eq. \ref{fomerGHZ} holds.
The proof is given by the following three steps:
\begin{enumerate}

\item When the quantum state is the single particle state $\left|a_i\right\rangle$, we have
$U \left|a_i\right\rangle \left|\varepsilon\right\rangle = 
\lambda_{a_ia_i} \left|a_i\right\rangle \left| \epsilon_{a_ia_i} \right\rangle$
(see Eq. \ref{single}).

\item Suppose that when the GHZ state is
$\left|0a_1a_2\cdots a_{n-1}\right\rangle$, the following equation holds,
\begin{align}
&U_0 \otimes U_1 \otimes \cdots \otimes U_{n-1} \left|0a_1a_2\cdots a_{n-1}\right\rangle
\left|\varepsilon\right\rangle_0 \left|\varepsilon\right\rangle_1 
\otimes \cdots \otimes
\left|\varepsilon\right\rangle_{n-1}  \notag \\
%%%%%%%%%%%%%%%%%%%%%%%%%%%%%%%%%%%%%%
& = \lambda_{00}	\lambda_{a_1a_1} \cdot \cdots \cdot \lambda_{a_{n-1}a_{n-1}}
\left|0a_1a_2\cdots a_{n-1}\right\rangle
\left| \epsilon_{00} \right\rangle 
\left| \epsilon_{a_1a_1} \right\rangle \otimes \cdots \otimes
\left| \epsilon_{a_{n-1}a_{n-1}} \right\rangle,
\end{align}

\item Then, when the state is $\left|0a_1a_2\cdots a_n\right\rangle$, we have
\begin{align}
&U_0 \otimes U_1 \otimes \cdots \otimes U_n \left|0a_1a_2\cdots a_n\right\rangle
\left|\varepsilon\right\rangle_0 \left|\varepsilon\right\rangle_1 
\otimes \cdots \otimes
\left|\varepsilon\right\rangle_n  \notag \\
%%%%%%%%%%%%%%%%%%%%%%%%%%%%%%%%%%%%%%
&= U_0 \otimes U_1 \otimes \cdots \otimes U_{n-1} \left|0a_1a_2\cdots a_{n-1}\right\rangle
\left|\varepsilon\right\rangle_0 \left|\varepsilon\right\rangle_1 
\otimes \cdots \otimes
\left|\varepsilon\right\rangle_{n-1} 
\otimes
U_n \left|a_n\right\rangle \left|\varepsilon\right\rangle_n 	
\notag \\
%%%%%%%%%%%%%%%%%%%%%%%%%%%%%%%%%%%%%%
& = \lambda_{00}	\lambda_{a_1a_1} \cdot \cdots \cdot \lambda_{a_{n-1}a_{n-1}}
\left|0a_1a_2\cdots a_{n-1}\right\rangle
\left| \epsilon_{00} \right\rangle 
\left| \epsilon_{a_1a_1} \right\rangle \otimes \cdots \otimes
\left| \epsilon_{a_{n-1}a_{n-1}} \right\rangle
\otimes \lambda_{a_na_n} \left|a_n\right\rangle \left|\epsilon_{a_na_n}\right\rangle
\notag \\
& = \lambda_{00}	\lambda_{a_1a_1} \cdot \cdots \cdot \lambda_{a_na_n}
\left|0a_1a_2\cdots a_n\right\rangle
\left| \epsilon_{00} \right\rangle 
\left| \epsilon_{a_1a_1} \right\rangle \otimes \cdots \otimes
\left| \epsilon_{a_na_n} \right\rangle,
\end{align}

\end{enumerate}

\begin{comment}

\subsubsection{Security under quantum channel noise}

Experiments in recent years show that the quantum bit error rate (QBER) caused by noise is 
between 1\% and 10\%, where the quantum channels used in the experiments 
include optical fibers and free-space \cite{WangS3762012,NauerthS752013,WangJY752013,
ValloneG11542015,BourgoinJP23262015,PirandolaS962015,LiaoSK1182017,LiaoSK5492017}.
However, the QBER caused by intercepting and measuring a decoy photon is 25\%.
That is, participants can detect the attacks of Eve
with the probability of 25\% using only one decoy photon.
Further, if Eve intercepts and measures $l$ decoy photons, 
the probability of detecting her attacks is $1-(3/4)^l$,
which will get closer and closer to 1 with the increase in $l$.
Obviously, the QBER caused by Eve's interference to
decoy photons is much higher than that caused by the noise.
Therefore, Eve cannot eavesdrop successfully under the cover of channel noise
as long as a threshold is set reasonably in advance 
according to the noise, which also shows the effectiveness of 
detecting eavesdropping with decoy photons.

\end{comment}

%-------------------------------------------------------------------------------------------

\subsection{Internal attacks}

Compared with external attackers, if the executors of a protocol are dishonest, 
they will pose a greater threat to the security of the protocol \cite{GaoF742007}.
In what follows we will analyze the internal attacks in detail from the following two aspects: 
one is that a dishonest participant tries to steal the secret data from another, 
the other is that TP attempts to steal the data from one or both participants.

\subsubsection{Attacks from TP}

Without loss of generality, we assume that TP manages to steal Alice's secret data.
Throughout our protocol, Alice's data $G_{a}^i$ is essentially encrypted by 
$K_{AB}^i$ and $K_{AC}^i$, in which the value of $K_{AB}^i$ is unknown to TP,
thus TP can guess $G_{a}^i$ with the successful probability of 1/2.
In this case, TP can guess Alice's data $X$ with the successful probability of $1/2^{\lceil N/n \rceil}$
(see the second prerequisite of our protocol),
where $1/2^{\lceil N/n \rceil}$ decreases with the increase in $\lceil N/n \rceil$.
Therefore, the probability that TP guesses Alice's secret data can be made
as small as $1/2^{\lceil N/2 \rceil}$ and as large as 1/2 
by varying the value of $n$.
Note that $N$ is usually a large number (see the first prerequisite of our protocol),
and the larger the value of $N$, the longer the bit length of its binary representation.
Even if the value of $N$ is small, Alice and Bob can increase it in many confidential ways. 
For example, they can agree in advance on a secret positive integer $M$ 
(or generate it as an additional key in the first step of the protocol),
whose value is confidential, and then they can calculate
$N \times M$, or $N^M$. In this way, just make sure that $\lceil N/n \rceil$ is big enough
so as to make $1/2^{\lceil N/n \rceil}$ small enough.
Of course, actual situations should be taken into account in the way to 
increase the value of $N$. After all, the larger the value of $N$, 
the more quantum resources the protocol consumes.

\subsubsection{Attacks from one dishonest participant}

Let us assume that Alice is dishonest and tries to steal Bob's data
because of the same role that two participants play.
Throughout our protocol, there are no qubits exchanged between Alice and Bob.
Therefore, only by intercepting the particles that Bob sends to TP
and measuring these particles after Bob tells TP the positions and bases of the decoy photons
can Alice have the chance to steal Bob's data.
In Step 2, Bob decides whether to flip the value of $G_b^i$ 
according to the value of $K_{AB}^i \oplus K_{BC}^i$;
the value of $K_{BC}^i$ is unknown to Alice.
In this case, Alice can guess the value of $G_b^i$ with the probability of 1/2.
Similar to the analysis above,
it is easy to see that Alice's probability of guessing Bob's data $Y$ is also $1/2^{\lceil N/n \rceil}$.
Additionally, Alice's attack will of course be detected 
as an external attacker during eavesdropping checking 
even if she intercepts Bob's particles and replaces them with false particles,
because at this time she has no idea of the positions and bases of the decoy photons.
Therefore, Alice's attacks will not succeed.

\subsubsection{Existing attack means against QPC}
\label{attack-means}

We now consider whether the attacks against the existing QPC protocols pose a threat to our protocol.
Let us start with a brief review of these attacks.
%(我们首先想要考虑我们最近提出的攻击手段。。。。接下来我们考虑其他手段)
At present, the existing attack means focus on internal attacks,
and each one is aimed at a specific protocol.
There is no universal attack means, which is, after all, difficult to propose because different protocols adopt 
different technologies, or algorithms, or both.
According to the difference of attackers, the existing attack means can be divided into two categories: 
one is the attack means of dishonest TP, the other is the attack means of dishonest participants (i.e., Alice and Bob).
The main means of TP is to steal the participants' secret data by fabricating the quantum states 
which are used as the information carriers in the protocol,
including the ones presented in Refs. \cite{YangYG1222013,TingX5632017,ChangY3312016,WangC11042013,GaoX5762018}.
The main means of the participants is that
one participant steals the secret data of another by intercepting the particles transmitted between him and TP,
including the ones presented in Refs. \cite{ZhangW1252013,WuW5862019,LiuXT8762013}.
Indeed, these attacks make many protocols insecure, that is, 
the protocols have information leakage problem. Fortunately, these attacks are invalid for our protocol.
In fact, each attack means is only used to attack a specific protocol, 
but not used to attack other protocols,
% (after all, it has not been proved to be effective against other protocols)
which is fundamentally due to the different quantum technologies and algorithms adopted by different protocols.
%Specifically, %下面让我们详细地探讨已经的攻击

Next, let us show in detail that the attack means mentioned above are invalid for our protocol.
Let us first analyze the attack means against the single-particle-based QPC protocols,
including the attack means proposed in Refs. \cite{WuWQ1722019,YangCW1282013}.
The success of the attack means in Ref. \cite{WuWQ1722019} is attributed to the fact that the protocol under
attack does not use decoy photon technology for eavesdropping checking,
while the attack means in Ref. \cite{YangCW1282013} can attack successfully due to
the two non-orthogonal bases as information carriers can be measured by a special basis.
These two attacks can not be used to attack our protocol, because our protocol uses decoy photon technology
for eavesdropping checking and uses entangled states as information carriers instead of single particles.
Let us then analyze the attack means against the entanglement-based QPC protocols,
including the attack means against the protocols using entanglement swapping \cite{ChenYT5332014,LiuW6222014},
the one against the protocols using unitary operations \cite{TingX5632017},
the ones against the protocol using one-way hash functions \cite{HeGP1462015,ZhangB14122015},
and the ones against the protocols using entanglement correlation 
but not using the above three technologies \cite{YangYG1222013,ChangY3312016,WuW5862019,WangC11042013}.
In the attack means of \cite{ChenYT5332014,LiuW6222014}, TP either infers the participants' data directly 
through the initial entangled states and the entangled states after entanglement swapping, 
or steals the participants' data by preparing fake entangled states before entanglement swapping.
Obviously,  these attacks are invalid for our protocol, because our protocol does not use entanglement swapping technology, 
and the quantum states used are prepared by participants rather than by TP.
The attack means in Refs. \cite{TingX5632017,HeGP1462015,ZhangB14122015} 
cannot attack our protocol since the encryption method 
adopted in the protocols they attack is different from that of our protocol.
Other attack means can be divided into two types: one is that TP attacks by preparing fake quantum states, 
which is obviously invalid for our protocol; the other is that the participants attack by the intercept-resend attack,
which is also invalid (see Sec. \ref{intercept-resend}).

\begin{comment}

The main means of TP is to fabricate the quantum states used in the protocol as the information carrier,
so as to steal the secret data of the participants.
The main attack means of the third party is to fabricate the quantum state which is used as the information carrier, 
and steal the secret data of the participants in this way.

\end{comment}

\section{Discussion}

In this section, we will first calculate qubit efficiency,
and then make a comparison between our protocol and some existing ones.

Qubit efficiency is defined as $ \eta_e = c/t$, where $c$
denotes the number of compared classical bits,
and $t$ the number of consumed particles, 
excluding the decoy photons and those consumed in the process of 
generating the keys using QKD \cite{JiZX1249112019}. 
In our protocol, two ($n+1$)-qubit GHZ states ($n \in \rm N_+$ and $2 \le n \le N$)
are used to achieve the comparison of $n$ bits of classical information
(i.e., the comparison of $G_a^i$ and $G_b^i$ requires two ($n+1$)-qubit GHZ states).
Therefore, the qubit efficiency of our protocol is $\frac {n}{2n+2}$.
Obviously, $\frac {n}{2n+2}$ increases with the increase in $n$,
hence we have 
\begin{align}
\label{qubitrange}
\frac 13 \le \frac {n}{2n+2} \le \frac{N}{2N+2} < \frac 12,
\end{align}
and
\begin{align}
\lim_{N \to + \infty} \frac {N}{2N+2} = \frac 12,
\end{align}
which means $\frac {N}{2N+2}$ will get closer and closer to 1/2 
with the increase in $N$.

We have shown that by adjusting the value of $n$, we can determine
the probability that TP can guess the participants' secret data ($1/2^{\lceil N/n \rceil}$),
the probability that a malicious participant can guess another's data,
and determine the qubit efficiency ($n/(2n+2)$).
Of course, the value of $n$ also determines
how many qubits that a ($n+1$)-qubit GHZ state contains,
thus determines the difficulty of preparing the state.
Generally, the more qubits a quantum state contains,
the more difficult it is to prepare and manipulate the quantum state,
with many challenges still remain in the
preparation and manipulation of multi-particle entanglement
\cite{DLeibfried43870682005,GaoWB652010,MonzT106132011,WangXL120262018}.
Fortunately, a series of significant progress has been made
in the preparation and manipulation of multi-particle GHZ states in recent years.
Recent research results show that 18-qubit entangled GHZ states
have been successfully prepared experimentally \cite{WangXL120262018}.

In Table \ref{Table}, we make the comparison between
our protocol and the existing ones proposed in Refs.
\cite{ChenXB2832010,LiuW512012,LiuW28412,ZhangWW532014,
LiJ532014,LiuW5162012,JiZX1652016,YeTY562017,LiC18158,
XuGA10042012,LiuWJ536,ZhaXW57122018,WangF59112016}.
In our protocol, the GHZ states are prepared by participants rather than by TP,
which to some extent, improves the security and efficiency of the protocol.
Indeed, in order to achieve the purpose of private comparison,
the participants will not prepare fake quantum states.
Nevertheless, in most existing QPC protocols,
the quantum states acting as information carriers are prepared by TP, 
in which case the protocols face more security risks (see \ref{attack-means}).
After all, the more work TP undertakes in the protocol, 
the more chance he has to counterfeit.
In addition, it is known that preparing and measuring devices for 
quantum states are necessary in QPC,
in which the prepared quantum states act as information carriers 
and the measurements are usually used to extract information contained in the quantum states.
In our protocol, except for the necessary devices for preparing quantum states, 
only single-particle measurement technology is used without any additional technology.
Therefore, our protocol has advantages over many existing protocols 
from security and device consumption.

%——————————————————————————————————table

\begin{table*}[h]
\begin{spacing}{1.0}
\centering
\setlength{\tabcolsep}{2pt}
\caption{Comparison between our protocol and some existing ones}
\label{Table}
\begin{tabular}{cccccccccccccccc}

\hline

Reference  
& \cite{JiZX1652016} & \cite{YeTY562017} & \cite{ChenXB2832010} & \cite{LiuW512012} & \cite{LiuW28412}
& \cite{ZhangWW532014} & \cite{LiJ532014}
& \cite{LiuW5162012} 
& \cite{LiC18158} & \cite{XuGA10042012}
& \cite{LiuWJ536} & \cite{ZhaXW57122018} & \cite{WangF59112016} 
& \makecell{Our protocol} \\

\hline

\makecell{QKD}
& $\surd$ & $\surd$ & $\times$ & $\surd$ & $\surd$ & $\surd$ & $\surd$ 
& $\surd$ & $\surd$ & $\surd$
& $\surd$ & $\surd$ & $\surd$ & $\surd$ \\

\noalign{\smallskip}\noalign{\smallskip}

\makecell{Unitary operation} 
& $\times$ & $\times$ & $\surd$ & $\times$ & $\surd$ & $\surd$ & $\times$
& $\surd$ & $\times$ 
& $\surd$ & $\surd$ & $\times$ & $\times$ & $\times$ \\

\noalign{\smallskip}\noalign{\smallskip}

\makecell{Entanglement swapping}
& $\times$ & $\times$ &$\times$ & $\surd$ & $\times$ & $\times$ & $\surd$
& $\times$ & $\surd$ 
& $\times$ & $\times$ & $\times$ & $\surd$ & $\times$ \\

\noalign{\smallskip}\noalign{\smallskip}

\makecell{Entanglement measurement}
& $\surd$ & $\surd$ &$\times$ & $\surd$ & $\times$ & $\times$ & $\surd$
& $\surd$ & $\surd$
& $\times$ & $\times$ & $\surd$ & $\surd$ & $\times$ \\

\noalign{\smallskip}\noalign{\smallskip}

\makecell{Qubit efficiency}
& $\frac 13$ & $\frac 25$ & $\frac 13$ & $\frac 13$ & $\frac 13$ & $\frac 13$ & $\frac 14$ & $\frac 14$
& $\frac 12$ & $\frac 12$ & $\frac 13$ & $\frac 25$
& $\frac 12$ & $\frac 13 \le \eta < \frac 12$ \\

\hline

\end{tabular}
\end{spacing}
\end{table*}

\section{Conclusion}

 %我们相信协议的安全性分析，尤其是对攻击手段的回顾，对于之后的QPC协议的设计和分析具有启发性和指导意义。
%纠缠测量攻击适用于。。。。

We have proposed a novel QPC protocol,
in which the entanglement correlation of the $(n+1)$-qubit GHZ state and bit-flipping
play key roles in the comparison of participants' secret data.
We have shown that TP and Eve cannot steal any useful information about
the participants' data, and that Alice and Bob cannot successfully steal each other's data.
Our protocol uses single-particle measurement technology to
extract information contained in quantum states,
which is easier to implement than entanglement measurements
under existing technical conditions.
The GHZ states are prepared by participants rather than by TP,
which makes it unnecessary for our protocol to
verify their authenticity. The participants can choose an appropriate 
value for $n$ according to actual situations.
That is, they can determine the value of $n$ based on 
how many qubits of the GHZ state can be prepared by their own devices.
Indeed, the larger the value of $n$, the higher the qubit efficiency,
but this also means the increase of the probability
that TP can guess the participants' data although the probability is very small.
Regardless of this, our protocol has some flexibility 
because of the selectivity that the participants have in the preparation of the GHZ states.

We wish that the algorithm adopted in our protocol and the security analysis, especially
the security proof against the entanglement-measurement attack,
can play an enlightening role in the design and security analysis of the GHZ-based QSMC protocols including QPC.
%我们不知道针对量子安全多方计算其他分支的攻击手段是否对我们的协议有效，这需要进一步的研究。

%-----------------------------------------------------------------------------------------------------


\begin{thebibliography}{99}
\addtolength{\itemsep}{-0.8 em}

\bibitem{ChoiSS812010}
Choi, S. S., Cha, S. H., and Tappert, C. C. (2010). 
A survey of binary similarity and distance measures. 
Journal of Systemics, Cybernetics and Informatics, 8(1), 43-48.

\bibitem{VinodP2009}
Vinod, P., Jaipur, R., Laxmi, V., and Gaur, M. (2009, March). 
Survey on malware detection methods. 
In Proceedings of the 3rd Hackers' Workshop on computer 
and internet security (IITKHACK'09) (pp. 74-79).

\bibitem{ZhangHG16102019}
H.-G. Zhang, Z.-X. Ji, H.-Z. Wang, W.-Q. Wu,
``Survey on quantum information security,'' China Commun., vol.16, no. 10, pp. 1-36, 2019.


\bibitem{JiaHY2841}
H.-Y. Jia, Q.-Y. Wen, T.-T. Song, and F. Gao, 
``Quantum protocol for millionaire problem,''
Opt. Commun., vol. 284, no. 1, pp. 545-549, 2011. 


\bibitem{YangYG422009}
Y.-G. Yang, and Q.-Y. Wen,
``An efficient two-party quantum private comparison protocol 
with decoy photons and two-photon entanglement,''
J. Phys. A: Math. Theor., vol. 42, no. 5, p. 055305, 2009.

\bibitem{JiZX1249112019}
Z.-X Ji, P.-R. Fan, H.-G. Zhang, and H.-Z. Wang,
``Several two-party protocols for quantum private comparison using entanglement and dense coding,'' 
Opt. Commun., vol. 495, no. 15, 2020.

\bibitem{ZXJi101007}
Z.-X. Ji, H.-G. Zhang, H.-Z. Wang, F.-S. Wu, J.-W. Jia, and W.-Q. Wu,  
``Quantum protocols for secure multi-party summation,''
Quantum Inf. Process., vol. 18, no. 6, p. 168, 2019.

\bibitem{ZhangHG58112015}
H.-G. Zhang, and X. J. Lai, et. al.,
``Survey on cyberspace security,'' Sci. China Inf. Sci., vol. 58, no. 11, pp. 1-43, 2015.

\bibitem{JiZX19112019}
Ji, Z. X., Fan, P. R., and Zhang, H. G. (2019). 
Entanglement swapping of Bell states and a special class
of Greenberger-Horne-Zeilinger states. arXiv preprint
arXiv:1911.09875.

%%%%%%%%%%%%%%%%

\bibitem{AlleaumeR5602014}
R. All\'eaume, et. al.,
``Using quantum key distribution for cryptographic purposes: a survey,''
Theor. Comput. Sci., vol. 560, pp. 62-81, 2014.

\bibitem{RennerR6012008}
R. Renner, ``Security of quantum key distribution,''
Int. J. Quantum Inf., vol. 6, no. 01, pp. 1-127, 2008.

\bibitem{GisinN7412002}
N. Gisin, et. al.,
``Quantum cryptography,'' Rev. Mod. Phys., vol. 74, no. 1, p. 145, 2002.

\bibitem{DengFG7012004}
F. G. Deng, and G. L. Long,
``Bidirectional quantum key distribution protocol with practical faint laser pulses,''
Phys. Rev. A, vol. 70, no. 1, p. 012311, 2004.

\bibitem{LinS8732013}
Lin, S. , Huang, C. , \& Liu, X. F. . (2013). Multi-user quantum key distribution based on bell states 
with mutual authentication. Physica Scripta, 87(3), 035008.

\bibitem{GuoY8142010}
Guo, Y., Shi, R., \& Zeng, G. . (2010). Secure networking quantum key distribution schemes with 
greenberger–horne–zeilinger states. Physica Scripta, 81(4), 045006.


\bibitem{JiZX1652016}
Z.-X. Ji, and T.-Y. Ye,
``Quantum private comparison of equal information based on 
highly entangled six-qubit genuine state,''
Commun. Theor. Phys., vol. 65, no. 6, p. 711, 2016.

\bibitem{YeTY562017}
T.-Y. Ye, and Z.-X. Ji,
``Two-party quantum private comparison with five-qubit entangled states,''
Int. J. Theor. Phys., vol. 56, no. 5, pp. 1517-1529, 2017.


%%%%%%%%%%%%%%%%%%%%%%%%


\bibitem{GaoF742007}
F. Gao, S.-J. Qin, Q.-Y. Wen, and F.-C. Zhu,
``A simple participant attack on the br\'{a}dler-du\v{s}ek protocol,''
Quantum Inf. \& Comput., vol. 7, no. 4, pp. 329-334, 2007.



%-------------------------



\bibitem{YangYG1222013}
Yang, Y., Xia, J., Jia, X., \& Zhang, H. (2013). 
Comment on quantum private comparison protocols with a semi-honest third party. 
Quantum Information Processing, 12(2), 877-885.

\bibitem{TingX5632017}
Ting, X., \& Tian-Yu, Y. (2017). Cryptanalysis and improvement for the quantum private comparison protocol 
based on triplet entangled state and single-particle measurement. International Journal of Theoretical Physics, 56(3), 771-780.

\bibitem{ChangY3312016}
Chang, Y., and Xu, C. X., et. al. (2016). 
Cryptanalysis and improvement of the multi-user QPCE protocol 
with semi-honest third party. Chin. Phys. L., 33(1), 010301.

\bibitem{WangC11042013}
C. Wang, G. Xu, and Y.-X. Yang,
``Cryptanalysis and improvements for the
quantum private comparison protocol using EPR pairs,''
Int. J. Quantum Inf., vol. 11, no. 04, p. 1350039, 2013.

\bibitem{GaoX5762018}
Gao, X. , Zhang, S. B. , Chang, Y. , Yang, F. , \& Zhang, Y. . (2018). 
Cryptanalysis of the quantum private comparison protocol based on the entanglement swapping 
between three-particle w-class state and bell state. International Journal of Theoretical Physics, 57(6), 1-7.

\bibitem{ZhangW1252013}
Zhang, W. W. , \& Zhang, K. J. . (2013). Cryptanalysis and improvement of the 
quantum private comparison protocol with semi-honest third party. , 12(5), 1981-1990.

\bibitem{WuW5862019}
Wu, W., Cai, Q., Wu, S., \& Zhang, H. (2019). Cryptanalysis and Improvement of Ye et al's 
Quantum Private Comparison Protocol. International Journal of Theoretical Physics, 58(6), 1854-1860.

\bibitem{LiuXT8762013}
Liu, X. T. , Zhao, J. J. , Wang, J. , \& Tang, C. J. . (2013).
Cryptanalysis of the secure quantum private comparison protocol. Physica Scripta, 87(6), 065004.

\bibitem{WuWQ1722019}
Wu, W. Q., Cai, Q. Y., Wu, S. M., \& Zhang, H. G. (2019). 
Cryptanalysis of he's quantum private comparison protocol and a new protocol. 
International Journal of Quantum Information, 17(2), 1950026.

\bibitem{YangCW1282013}
Yang, C. W., Kao, S. H., \& Hwang, T. (2013).
Comment on "efficient and feasible quantum private comparison of equality against the collective amplitude damping noise". 
Quantum Information Processing, 12(8), 2871-2875.

\bibitem{ChenYT5332014}
Chen, Y. T., \& Hwang, T. (2014). Comment on the "quantum private comparison protocol based on bell entangled states". 
International Journal of Theoretical Physics, 53(3), 837-840.

\bibitem{LiuW6222014}
Liu, W., Liu, C., Chen, H., Li, Z., \& Liu, Z. (2014). Cryptanalysis and Improvement of Quantum Private Comparison 
Protocol Based on Bell Entangled States. Communications in Theoretical Physics, 62(2), 210-214.

\bibitem{HeGP1462015}
He, G. P. (2015). Comment on Quantum private comparison of equality protocol without a third party. 
Quantum Information Processing, 14(6), 2301-2305.

\bibitem{ZhangB14122015}
Zhang, B., Liu, X. , Wang, J., \& Tang, C. (2015). Cryptanalysis and improvement of 
quantum private comparison of equality protocol without a third party. Quantum Information Processing, 14(12), 4593-4600.



\bibitem{DLeibfried43870682005}
D. Leibfried, et. al.,
``Creation of a six-atom `Schr\"odinger cat' state,'' Nature, vol. 438, no. 7068, p. 639, 2005.

\bibitem{GaoWB652010}
W.-B. Gao, C.-Y. Lu, X.-C. Yao, P. Xu, O. G\"uhne, A. Goebel,\dots and J.-W. Pan,
``Experimental demonstration of a hyper-entangled ten-qubit Schr\"odinger cat state,''
Nat. phys., vol. 6, no. 5, p. 331, 2010.

\bibitem{MonzT106132011}
T. Monz, P. Schindler, J. T. Barreiro, M. Chwalla, D. Nigg, W. A. Coish,\dots and R. Blatt, 
``14-qubit entanglement: Creation and coherence,''
Phys. Rev. Lett., vol. 106, no. 13, p. 130506, 2011.

\bibitem{WangXL120262018}
X.-L. Wang, Y.-H. Luo, H.-L. Huang, M.-C. Chen, Z.-E. Su, C. Liu, \dots and J. Zhang,
``18-qubit entanglement with six photons' three degrees of freedom,''
Phys. Rev. Lett., vol. 120, no. 26, p. 260502, 2018.





%-------------------------下面不变

\bibitem{ChenXB2832010}
X.-B. Chen, G. Xu, X.-X. Niu, Q.-Y. Wen, and Y.-X. Yang,
``An efficient protocol for the private comparison of equal information
based on the triplet entangled state and single-particle measurement,''
Opt. Commun., vol. 283, no. 7, pp. 1561-1565, 2010.

\bibitem{LiuW512012}
W. Liu, and Y.-B. Wang,
``Quantum private comparison based on GHZ entangled states,''
Int. J. Theor. Phys., vol. 51, no. 11, pp. 3596-3604, 2012.

\bibitem{LiuW28412}
W. Liu, Y.-B. Wang, and Z.-T. Jiang,
``An efficient protocol for the quantum private comparison of equality with W state,"
Opt. Commun., vol. 284, no. 12, pp. 3160-3163, 2011. 


\bibitem{ZhangWW532014}
W.-W. Zhang, D. Li, and Y.-B. Li,
``Quantum private comparison protocol with W States,''
Int. J. Theor. Phys., vol. 53, no. 5, pp. 1723-1729, 2014.

\bibitem{LiJ532014}
J. Li, H.-F. Zhou, L. Jia, and T.-T.  Zhang,
``An efficient protocol for the private comparison of equal information based on 
four-particle entangled W state and Bell entangled states swapping,'' 
Int. J. Theor. Phys., vol. 53, no. 7, pp. 2167-2176, 2014.

\bibitem{LiuW5162012}
W. Liu, Y.-B. Wang, Z.-T. Jiang, Y.-Z. Cao, and W. Cui,
``New quantum private comparison protocol using $\upchi$-type state,'' 
Int. J. Theor. Phys., vol. 51, no. 6, pp. 1953-1960, 2012.

\bibitem{LiC18158}
C. Li,  X. Chen, H. Li,  Y. Yang, and J. Li,
``Efficient quantum private comparison protocol based on 
the entanglement swapping between four-qubit cluster state and extended Bell state,''
Quantum Inf. Process., vol. 18, no. 5, p. 158, 2019.


\bibitem{XuGA10042012}
G.-A. Xu, X.-B. Chen, Z.-H. Wei, M.-J. Li, and Y.-X. Yang,
``An efficient protocol for the quantum private comparison of 
equality with a four-qubit cluster state,''
Int. J. Quantum Inf., vol. 10, no. 04, p. 1250045, 2012. 

\bibitem{LiuWJ536}
W.-J. Liu, C. Liu, H.-B. Wang, J.-F. Liu, F. Wang, and X.-M. Yuan,
``Secure quantum private comparison of equality based on asymmetric W state,''
Int. J. Theor. Phys., vol. 53, no. 6, pp. 1804-1813, 2014. 

\bibitem{ZhaXW57122018}
X.-W. Zha, X.-Y. Yu, Y. Cao, and S.-K. Wang,
``Quantum Private Comparison Protocol with Five-Particle Cluster States,''
Int. J. Theor. Phys., vol. 57, no. 12, pp. 3874-3881, 2018.

\bibitem{WangF59112016}
Wang, F., Luo, M., Li, H., Qu, Z., and X. Wang, 
Quantum private comparison based on quantum dense coding. 
Sci. China Inf. Sci., vol. 59, no. 11, p. 112501, 2016.




\begin{comment}



\bibitem{SunZW12042012}
Z.-W. Sun, and D.-Y. Long, 
``Cryptanalysis of the efficient two-party quantum private comparison protocol 
with decoy photons and two-photon entanglement.,'' 
arXiv preprint arXiv:1204.4587. 2012.



%————抗噪声

\bibitem{WangS3762012}
S. Wang, W. Chen, J.-F. Guo, Z.-Q. Yin, H.-W. Li, Z. Zhou,\dots and Z. F. Han, 
``2 GHz clock quantum key distribution over 260 km of standard telecom fiber,''
Opt. lett., vol. 37, no. 6, pp. 1008-1010, 2012.

\bibitem{NauerthS752013}
S. Nauerth, et. al., 
``Air-to-ground quantum communication,'' Nat. Photonics, vol. 7, no. 5, p. 382, 2013.

\bibitem{WangJY752013}
J.-Y. Wang, et. al.,  
``Direct and full-scale experimental verifications towards ground-satellite 
quantum key distribution,'' Nat. Photonics, vol. 7, no. 5, p. 387, 2013.

\bibitem{ValloneG11542015}
G. Vallone, D. Bacco, D. Dequal, S. Gaiarin, V. Luceri, G. Bianco,
and P. Villoresi, ``Experimental satellite quantum communications,''
Phys. Rev. Lett., vol. 115, no. 4, p. 040502, 2015.

\bibitem{BourgoinJP23262015}
J. P. Bourgoin, et. al., ``Free-space quantum key distribution to a moving receiver,''
Opt. Express, vol. 23, no. 26, pp. 33437-33447, 2015.

\bibitem{PirandolaS962015}
S. Pirandola, et. al.,
``High-rate measurement-device-independent quantum cryptography,''
Nat. Photonics, vol. 9, no. 6, p. 397, 2015.

\bibitem{LiaoSK1182017}
S.-K. Liao, et. al.,
``Long-distance free-space quantum key distribution in daylight towards 
inter-satellite communication,'' Nat. Photonics, vol. 11, no. 8, p. 509, 2017.

\bibitem{LiaoSK5492017}
S.-K. Liao, et. al.,
``Satellite-to-ground quantum key distribution,'' Nature, vol. 549, no. 7670, p. 43, 2017.
%————抗噪声

%------------------------------------------------------------------------------------------------------------QPC


\bibitem{LiuWJ3052013}
W.-J. Liu, C. Liu,  H. Wang, and T. Jia, ``Quantum private comparison: a review,"
IETE Tech. Rev., vol. 30, no. 5, pp. 439-445, 2013.

\bibitem{ZXJi72019}
Z.-X. Ji, H.-G. Zhang, and H.-Z. Wang,
``Quantum Private Comparison Protocols With a Number of Multi-Particle Entangled States,''
IEEE Access, vol. 7, pp. 44613-44621, 2019.

\bibitem{YeTY6092017}
T.-Y. Ye, , and Z.-X. Ji, 
``Multi-user quantum private comparison with scattered preparation 
and one-way convergent transmission of quantum states,''
Sci. China Phys. Mech., vol. 60, no. 9, p. 090312, 2017.

\bibitem{LinJ1322014}
J. Lin, C.-W. Yang, and T. Hwang,
``Quantum private comparison of equality protocol without a third party,''
Quantum Inf. Process., vol. 13, no. 2, pp. 239-247, 2014.

\bibitem{ZhangB14122015}
B. Zhang, X.-T. Liu, J. Wang, and C.-J. Tang,
``Cryptanalysis and improvement of quantum private comparison of equality protocol 
without a third party,'' Quantum Inf. Process., vol. 14, pp. 4593-4600, 2015.

\bibitem{HeGP1462015}
G.-P. He,
``Comment on `Quantum private comparison of equality protocol without a third party',''
Quantum Inf. Process., vol. 14, no. 6, pp. 2301-2305, 2015.



\bibitem{JiZX162017}
Z.-X. Ji, and T.-Y. Ye,  
``Multi-party quantum private comparison based on the entanglement swapping 
of d-level cat states and d-level Bell states,''
Quantum Inf. Process., vol. 16, no. 7, p. 177, 2017.

\bibitem{LinJ2842011}
J. Lin, H.-Y. Tseng, and T. Hwang,
``Intercept-resend attacks on Chen et al.'s 
quantum private comparison protocol and the improvements,''
Opt. Commun., vol. 284, no. 9, pp. 2412-2414, 2011.

\bibitem{TXu563}
T. Xu, et. al., ``Cryptanalysis and improvement for the 
quantum private comparison protocol based on triplet entangled state 
and single-particle measurement,''
Int. J. Theor. Phys., vol. 56, no. 3, pp. 771-780, 2017.

\bibitem{LoHK5621997}
H. K. Lo, ``Insecurity of quantum secure computations,'' 
Phys. Rev. A, vol. 56, no. 2, p. 1154, 1997.

%----------------------------------------------------------------------------------------------------


\end{comment}










\end{thebibliography}
\end{document}